\documentclass[]{aastex62}

\accepted{November 5, 2018}

\submitjournal{ApJ}

\usepackage{xcolor}
\usepackage{graphicx}
\usepackage{pgfplots}
\pgfplotsset{compat=1.13}
\usepgfplotslibrary{polar}

\usepackage{savesym}
\savesymbol{tablenum}
\usepackage{siunitx}
\restoresymbol{SIX}{tablenum}

\usepackage{verbatim}
\usepackage{multirow}
\usepackage{textgreek}

\newcommand*\chem[1]{\ensuremath{\mathrm{#1}}}

\newcommand{\rev}[1]{#1} 
\newcommand{\revtwo}[1]{#1} 

\DeclareSIUnit \parsec {pc}
\DeclareSIUnit \solarmass {\mbox{M$_\sun$}}
\DeclareSIUnit \solarluminosity {\mbox{L$_\sun$}}
\DeclareSIUnit \solarradius {\mbox{R$_\sun$}}
\DeclareSIUnit \milliarcsecond {mas}
\DeclareSIUnit \au {au}
\DeclareSIUnit \year {yr}
\DeclareSIUnit \hourangle {\!\textsuperscript{h}}
\DeclareSIUnit \minuteangle {\!\textsuperscript{m}}
\DeclareSIUnit \secondangle {\!\textsuperscript{s}}
\begin{document}

\title{Probing the Inner Disk Emission of the Herbig Ae Stars HD~163296 and HD~190073}

\author[0000-0001-5980-0246]{Benjamin R. Setterholm}
\affiliation{University of Michigan, Astronomy Department, Ann Arbor, MI 48109, USA}
\email{bensett@umich.edu}

\author[0000-0002-3380-3307]{John D. Monnier}
\affiliation{University of Michigan, Astronomy Department, Ann Arbor, MI 48109, USA}

\author[0000-0001-9764-2357]{Claire L. Davies}
\affiliation{University of Exeter, Department of Physics and Astronomy, 
Exeter, Devon EX4 4QL, UK}

\author[0000-0002-0911-9505]{Alexander Kreplin}
\affiliation{University of Exeter, Department of Physics and Astronomy, 
Exeter, Devon EX4 4QL, UK}

\author[0000-0001-6017-8773]{Stefan Kraus}
\affiliation{University of Exeter, Department of Physics and Astronomy, 
Exeter, Devon EX4 4QL, UK}

\author[0000-0002-5074-1128]{Fabien Baron}
\affiliation{Georgia State University, Department of Physics and Astronomy, Atlanta, GA 30302, USA}

\author[0000-0002-1327-9659]{Alicia Aarnio}
\affiliation{University of Colorado Boulder, Boulder, CO 80303, USA}

\author{Jean-Philippe Berger}
\affiliation{Universit\'e Grenoble Alpes, CNRS, IPAG, 38000 Grenoble, France}

\author[0000-0002-3950-5386]{Nuria Calvet}
\affiliation{University of Michigan, Astronomy Department, Ann Arbor, MI 48109, USA}

\author{Michel Cur\'e}
\affiliation{Universidad de Valpara\'{\i}so, 
Instituto de F\'{\i}sica y Astronom\'{\i}a, 
Valpara\'{\i}so, Chile}

\author{Samer Kanaan}
\affiliation{Universidad de Valpara\'{\i}so, 
Instituto de F\'{\i}sica y Astronom\'{\i}a, 
Valpara\'{\i}so, Chile}

\author{Brian Kloppenborg}
\affiliation{Georgia State University, Department of Physics and Astronomy, Atlanta, GA 30302, USA}

\author{Jean-Baptiste Le Bouquin}
\affiliation{University of Michigan, Astronomy Department, Ann Arbor, MI 48109, USA}
\affiliation{Universit\'e Grenoble Alpes, CNRS, IPAG, 38000 Grenoble, France}

\author[0000-0003-0447-5866]{Rafael Millan-Gabet} 
\affiliation{California Institute of Technology, Infrared Processing and Analysis Center, Pasadena, CA 91125, USA}

\author{Adam E. Rubinstein}
\affiliation{University of Michigan, Astronomy Department, Ann Arbor, MI 48109, USA}

\author[0000-0003-1799-1755]{Michael L. Sitko}
\affiliation{University of Cincinnati, Department of Physics, Cincinnati, OH 45221, USA}
\affiliation{Space Science Institute, Center for Extrasolar Planetary Systems, 
Boulder, CO 80301, USA}

\author{Judit Sturmann}
\affiliation{The CHARA Array of Georgia State University, Mount Wilson Observatory, Mount Wilson, CA 91023, USA}

\author[0000-0002-0114-7915]{Theo A. ten Brummelaar}
\affiliation{The CHARA Array of Georgia State University, Mount Wilson Observatory, Mount Wilson, CA 91023, USA}

\author{Yamina Touhami}
\affiliation{Georgia State University, Department of Physics and Astronomy, Atlanta, GA 30302, USA}


\begin{abstract}

The physical processes occurring within the inner few astronomical units of proto-planetary disks surrounding Herbig Ae stars are crucial to setting the environment in which the outer planet-forming disk evolves and put critical constraints on the processes of accretion and planet migration.  We present the most complete published sample of high angular resolution H- and K-band observations of the stars \object{HD~163296} and \object{HD~190073}, including 30 previously unpublished nights of observations of the former and 45 nights of the latter with the CHARA long-baseline interferometer, in addition to archival VLTI data. We confirm previous observations suggesting significant near-infrared emission originates within the putative dust evaporation front of HD~163296 and show this is the case for HD~190073 as well.  The H- and K-band sizes are the same within \SI{3 \pm 3}{\percent} for HD~163296 and within \SI{6 \pm 10}{\percent} for HD~190073.  The radial surface brightness profiles for both disks are remarkably Gaussian-like  with little or no sign of the sharp edge expected for a dust evaporation front.
Coupled with spectral energy distribution analysis, our direct measurements of the stellar flux component at H and K bands suggest that \rev{HD~190073 is much younger ($<$\SI{400}{\kilo\year}) and more massive ($\sim$\SI{5.6}{\solarmass}) than previously thought, mainly as a consequence of the new {\it Gaia}  distance (\SI{891}{\parsec}).}

\end{abstract}


\section{Introduction} \label{sec:intro}

Herbig Ae (HAe) stars are a class of intermediate mass (\SIrange[range-units = single, range-phrase = --]{1.5}{5}{\solarmass}) pre-main-sequence stellar objects characterized by strong excess emission in the near-infrared (NIR) and millimeter wavelengths, typically peaking around \SI{3}{\micro\meter}.  Due to their young age and high luminosity, HAe stars are the ideal targets for observing accretion and planet formation processes {\em in situ}.  The bulk of the NIR excess originates from within the inner few astronomical units surrounding these stars.  Since even the closest HAe systems are located at distances exceeding \SI{100}{\parsec}, the milliarcsecond resolution required to observe features on \si{\au} scales is far beyond the capabilities of even the largest solitary optical telescopes.  However, advancements made in long-baseline optical interferometry over the course of the last two decades have made it possible to probe these spatial scales.

Early HAe investigations of \cite{Hillenbrand_1992}, limited only to spectral energy distribution (SED) measurements, were able to reproduce the photometric NIR excess measurements successfully by assuming a flat, optically-thick accretion disk that extends down to a few stellar radii.  This theoretical picture, however, was \emph{not confirmed} by the early infrared interferometry observations by \citet{MillanGabet_1999} of \object{AB~Aurigae} with the Infrared Optical Telescope Array, where they found NIR disk sizes many times larger than predicted by optically-thick, geometrically thin disk models. \citet{Natta_2001} and \citet{Dullemond_2001} put forward a new inner disk model where the star is surrounded by an optically-thin cavity with a ``puffed-up'' inner rim wall located at the radius where the equilibrium temperature is equal to the dust's characteristic sublimation temperature.  Additional interferometer measurements by \citet{MillanGabet_2001} along with Keck aperture masking observations of \object{LkH$\alpha$~101} \citep{Tuthill_2001} supported the idea that the bulk of the NIR disk emission comes from a ring located at the dust sublimation radius.  With these results, \citet{Monnier_2002} published the first ``size-luminosity diagram'' finding dust sublimation temperatures between \SIrange[range-units = single, range-phrase = --]{1500}{2000}{\kelvin}, broadly consistent across the sample.  The similarity between this characteristic temperature at $\sim$\SI{1800}{\kelvin} and known silicate sublimation temperatures indicated that the NIR emission is tracing a silicate dust sublimation rim.  \citet{Dullemond_2010} \revtwo{presents} a comprehensive overview of the theoretical and observational picture for the inner disks of young stellar objects (YSOs) that is still largely up-to-date.

The first sub-milliarcsecond interferometric observations (\citealt{Tannirkulam_2008}) of HAe stars \object{HD~163296} and \object{AB Aurigae} unexpectedly discovered that a significant fraction of the flux responsible for the NIR excess originated from well-within the supposed dust sublimation radius.  Moreover, they found it was unlikely that the distribution had a sharp edge
with an illuminated inner rim. Subsequent observations (for example, see \citealt{Benisty_2010} and \citealt{Lazareff_2017}) have confirmed this basic result across a large number of HAe stars, in contradiction to the prevailing theory of HAe inner disk structure. Several phenomena have been proposed to explain this large amount of inner emission, including emission from refractory dust grains (\citealt{Benisty_2010}) and optically thick free-free/bound-free emission from a hot accreting gas (\citealt{Kraus_2008}; \citealt{Tannirkulam_2008}).  Further multi-wavelength measurements of more HAe objects are required to determine the relevant mechanisms dominating the emission.

In this paper, we conduct a multi-wavelength interferometric study in H and K bands for two HAe objects: \object{HD~163296} (MWC~275) and \object{HD~190073} (V1295~Aql).  A list of basic stellar properties and NIR photometry from the literature for these two objects is reproduced in Table \ref{tab:basicprop}\rev{, where we adopt photosphere temperatures based solely on the measured spectral type, as literature temperature estimates range on the order of a few \revtwo{hundred} Kelvin for both objects}.  Our \rev{interferometric} data are collected on longer baselines with more complete $(u, v)$ coverage than is currently available for HAe stars in the literature.  The higher angular resolution better constrain the orientation and radial distribution of the material producing the emission than previous work while the multi-wavelength data probe the mechanism producing the mysterious interior emission.  We analyze the measurements by fitting simple geometrical models to the brightness distribution; our goal in this paper is to characterize the sizes and general profiles of these stars rather than try to model small scale details to which our data is not adequately sensitive.  We then validate our interferometric model fitting by comparing to the object SEDs and compute new luminosity, mass, and age estimates of these YSOs.  Finally, we speculate on the physical origin of the interior NIR excess emission.

\begin{deluxetable}{lcc}
    \tablecaption{Literature Stellar Properties and Photometry of Target Sources}
    \tablehead{\colhead{Property} & \colhead{\object{HD~163296}} & \colhead{\object{HD~190073}}}
    \startdata
        $\alpha$ (J2000)   &   \SI{17}{\hourangle}\,\SI{56}{\minuteangle}\,\SI{21.29}{\secondangle}
                            &   \SI{20}{\hourangle}\,\SI{03}{\minuteangle}\,\SI{02.51}{\secondangle} \\
        $\delta$ (J2000)   &   \ang[angle-symbol-over-decimal]{-21; 57; 21.87}
                            &   \ang[retain-explicit-plus,angle-symbol-over-decimal]{+05; 44; 16.66} \\[4pt]
        Spectral Type\tablenotemark{a}   &   A1 Vepv   &   A2 IVev  \\
        $T_\text{eff}$\tablenotemark{b}     &   \rev{\SI{9230}{\kelvin}}   &   \rev{\SI{8970}{\kelvin}} \\
        Distance\tablenotemark{c}           &   \SI{101.5 \pm 1.2}{\parsec}   &   \SI{891 \pm 53}{\parsec} \\
        Luminosity\tablenotemark{d}         &   \SI{28}{\solarluminosity}
                            &   \SI{780}{\solarluminosity} \\[4pt]
 $V$ mag\tablenotemark{e}   &   \num{6.84 \pm 0.06}   &   \num{7.79 \pm 0.06} \\
        $H$ mag\tablenotemark{e}   &   \num{5.48 \pm 0.07}   &   \num{6.61 \pm 0.07} \\
        $K$ mag\tablenotemark{e}   &   \num{4.59 \pm 0.08}   &   \num{5.75 \pm 0.08} \\
    \enddata
    \tablerefs{\textsuperscript{a}\cite{Mora_2001};\rev{ \textsuperscript{b}\cite{Kenyon_1995}, based on reported spectral type;} \textsuperscript{c}\cite{Gaia_DR2}; \textsuperscript{d}\cite{Monnier_2006}, rescaled to Gaia DR2 distances; \textsuperscript{e}\cite{Tannirkulam_2008}}
    \label{tab:basicprop}
\end{deluxetable}

\section{Observations and Data Reduction} \label{sec:obsdata}

In this section, we describe and present the most complete published sample of broadband H and K band long-baseline interferometric data collected of the HAe stars \object{HD~163296} and \object{HD~190073} by combining interferometer observations conducted at various facilities.  We will provide a copy of the calibrated data used in our modeling using the OI-FITS format \citep{Pauls_2005} and uploaded to the Optical interferometry Database (OiDb) \citep{Haubois_2014} developed by the Jean-Marie Mariotti Center (Grenoble, France). 

\subsection{CHARA Interferometric Data} \label{sec:chara}

New observations were conducted at the Center for High Angular Resolution Astronomy (CHARA) interferometer \citep{ten_Brummelaar_2005}, located on Mt.\@ Wilson, California, with baselines of up to \num{330} meters and at various orientations.  These data together yield a maximum nominal angular resolution of $\lambda/2B = \num{0.51}$ milliarcseconds (\si{\milliarcsecond}) in H band and \SI{0.67}{\milliarcsecond} in K band, where $\lambda$ is the filter central wavelength and $B$ is the longest baseline length measured.

We used the CHARA 2--telescope ``Classic'' beam-combiner instrument (\citealt{ten_Brummelaar_2013}) to collect broad H-band ($\lambda_{\text{eff}} = \SI{1.673}{\micro\meter}$, $\Delta\lambda = \SI{0.304}{\micro\meter}$) and K-band ($\lambda_{\text{eff}} = \SI{2.133}{\micro\meter}$, $\Delta\lambda = \SI{0.350}{\micro\meter}$) squared visibility ($\mathcal{V}^2$) measurements of our target stars between July 2004 and July 2010.  We supplement the eight nights of HD~163296 Classic observations previously published in \cite{Tannirkulam_2008} with 8 additional nights of data.  We also present sixteen new, formerly unpublished nights of observations of HD~190073.  A summary of the Classic measurements is given in Table~\ref{tab:classic}.

\begin{deluxetable}{rrlcclh}
    \tablecaption{CHARA/Classic Observations}
    \tablehead{\colhead{Target} & \colhead{Date} & \colhead{Configuration} & \colhead{Band} & \colhead{n.\,$V^2$} & \colhead{Calibrator(s)} & \nocolhead{Notes}}
    \startdata
    HD~163296 & 2004-07-09   &   S1-W1   &   K   &   2   &   \object{HD 164031}, \object{HD 163955}   &   \nodata   \\
    &   2005-07-20   &   W1-W2   &   K   &   2   &   \object{HD 164031}, \object{HD 163955}   &   obslog says 3 sets   \\
    &   2005-07-22   &   W1-W2   &   K   &   3   &   \object{HD 164031}, \object{HD 163955}   &   \nodata   \\
    &   2005-07-26   &   W1-W2   &   K   &   4   &   \object{HD 164031}, \object{HD 163955}   &   \nodata   \\
    &   2006-06-22   &   S2-W2   &   K   &   5   &   \object{HD 164031}, \object{HD 163955}   &   obslog says 3 sets   \\
    &   2006-06-23   &   W1-E1   &   K   &   2   &   \object{HD 163955}, \object{HD 164031}   &   obslog says 1 set   \\
    &      &   S2-W2   &   K   &   1   &   \object{HD 163955}, \object{HD 164031}   &   \nodata   \\
    &   2006-08-22   &   S2-E2   &   K   &   1   &   \object{HD 166295}   &   \nodata   \\
    &   2006-08-23   &   S2-E2   &   K   &   3   &   \object{HD 166295}, \object{HD 164031}, \object{HD 163955}   &   \nodata   \\
    &   2007-06-15   &   S2-W1   &   K   &   5   &   \object{HD 161023}, \object{HD 162255}   &   \nodata   \\
    &   2007-06-16   &   S2-W1   &   K   &   1   &   \object{HD 164031}   &   \nodata   \\
    &   2007-06-17   &   S2-W1   &   K   &   5   &   \object{HD 156365}, \object{HD 164031}   &   \nodata   \\
    &   2008-06-12   &   W1-W2   &   K   &   5   &   \object{HD 164031}, \object{HD 156365}   &   \nodata   \\
    &   2008-06-15   &   W1-W2   &   H   &   3   &   \object{HD 161023}, \object{HD 162255}   &   \nodata   \\
    &      &   S2-W1   &   H   &   2   &   \object{HD 161023}, \object{HD 162255}   &   \nodata   \\
    &   2008-06-16   &   S2-W1   &   H   &   1   &   \object{HD 161023}, \object{HD 162255}   &   \nodata   \\
    &   2009-06-24   &   S2-E2   &   H   &   3   &   \object{HD 156365}, \object{HD 164031}   &   \nodata   \\
    &   2009-06-25   &   S2-E2   &   H   &   2   &   \object{HD 156365}, \object{HD 164031}   &   obslog says 3 sets   \\ \hline
    HD~190073   &   2007-06-08     &   W1-W2   &   K   &   6   &   \object{HD 187923}, \object{HD 193556}    &   \nodata \\
    &   2007-06-09   &   W1-W2   &   K   &   6   &   \object{HD 187923}, \object{HD 193556}    &   \nodata \\
    &   2008-06-14   &   W1-W2   &   H   &   3   &   \object{HD 187923}, \object{HD 193556}    &   \nodata \\
    &   2008-06-17   &   W1-S2   &   K   &   1   &   \object{HD 187923}, \object{HD 193556}    &   clouds, large count fluctuations \\
    &      &   W1-E2   &   K   &   1   &   \object{HD 187923}                                   &   obslog says not reduced \\
    &   2008-06-18   &   W1-S2   &   K   &   2   &   \object{HD 187923}, \object{HD 193556}    &   \nodata\\
    &   2009-06-19   &   W1-S2   &   K   &   2   &   \object{HD 193556}                        &   \nodata \\
    &   2009-06-20   &   W1-S2   &   K   &   2   &   \object{HD 187923}, \object{HD 193556}    &   obslog says 1 set \\
    &   2009-06-22   &   W2-E2   &   K   &   5   &   \object{HD 183303}                        &   obslog says 4 sets \\
    &      &   E1-S1   &   K   &   1   &   \object{HD 183303}                                   &   not in obslog \\
    &   2009-06-24   &   E2-S2   &   H   &   3   &   \object{HD 183303}, \object{HD 183936}    &   obslog says 2 sets \\
    &   2009-06-25   &   E2-S2   &   K   &   5   &   \object{HD 183303}                        &   \nodata \\
    &   2009-10-31   &   W2-S2   &   K   &   5   &   \object{HD 183936}                        &   \nodata \\
    &   2010-06-14   &   W2-S1   &   K   &   2   &   \object{HD 150366}, \object{HD 188385}    &   \nodata \\
    &      &   E2-S1   &   K   &   3   &   \object{HD 187897}, \object{HD 189509}    &   \nodata \\
    &   2010-06-15   &   W1-E2   &   K   &   1   &   \object{HD 189509}                        &   \nodata \\
    &   2010-06-16   &   E1-S1   &   K   &   5   &   \object{HD 177305}, \object{HD 177332}, \object{HD 188385}, \object{HD 189509}    &   obslog says 4 sets \\
    &   2010-06-17   &   W1-S1   &   K   &   5   &   \object{HD 141597}, \object{HD 188385}, \object{HD 206660}   &   \nodata \\
    &   2010-07-22   &   S1-E1   &   K   &   3   &   \object{HD 188385}, \object{HD 189509}    &   \nodata \\
    \enddata
    \tablenotetext{}{The ``n.\@ $V^2$'' column indicates the number of squared visibility measurements collected per the particular observation.}
    \label{tab:classic}
\end{deluxetable}

Squared visibility and closure phase measurements were also obtained with the 3--telescope CLIMB instrument (\citealt{ten_Brummelaar_2013}) between July 2010 and June 2014.  Broad H-band ($\lambda_{\text{eff}} = \SI{1.673}{\micro\meter}$, $\Delta\lambda = \SI{0.274}{\micro\meter}$) and K-band ($\lambda_{\text{eff}} = \SI{2.133}{\micro\meter}$, $\Delta\lambda = \SI{0.350}{\micro\meter}$) observations were collected over the span of 22 nights for HD~163296 and 29 nights for HD~190073; a summary of these measurements is given in Table~\ref{tab:climb}.

\begin{deluxetable}{rrlccclh}
    \tablecaption{CHARA/CLIMB Observations}
    \tablehead{\colhead{Target} & \colhead{Date} & \colhead{Configuration} & \colhead{Band} & \colhead{n.\,$V^2$} & \colhead{n.\,$T^3$} & \colhead{Calibrator(s)} & \nocolhead{Notes}}
    \startdata
    HD~163296   &   2010-07-11   &   S1-S2-W2   &   K   &   11      &   1       &   \object{HD 164031}   &   \nodata \\
    &   2010-07-12   &   W1-E1-E2   &   K   &   7       &   1       &   \object{HD 164031}   &   \nodata \\
    &   2010-07-13   &   E1-E2      &   K   &   4       &   0       &   \object{HD 156365}   &   \nodata \\
    &   2010-07-14   &   W1-E1-E2	&   K   &   7       &   1       &   \object{HD 159743}, \object{HD 164104}   &   \nodata \\
    &   2010-07-15   &   W1-E1-E2   &   K   &   14      &   2       &   \object{HD 159743}, \object{HD 170657}   &   \nodata \\
    &   2011-06-15   &   W1-W2-E1   &   K   &   14      &   2       &   \object{HD 162255}, \object{HD 170680}   &   \nodata \\
    &   2011-06-20   &   W1-W2-E1   &   K   &   14      &   2       &   \object{HD 162255}, \object{HD 163955}   &   \nodata \\
    &   2011-06-23   &   S1-W1-W2   &   K   &   28      &   4       &   \object{HD 162255}, \object{HD 163955}   &   \nodata \\
    &   2011-06-24   &   S1-W2-E2   &   K   &   35      &   5       &   \object{HD 162255}, \object{HD 163955}   &   \nodata \\
    &   2011-06-27   &   S1-W1-E2   &   K   &   21      &   3       &   \object{HD 163955}   &   \nodata \\
    &   2011-06-30   &   W2-E2      &   K   &   3       &   0       &   \object{HD 163955}   &   obslog says this is S2-W2 \\
    &   2011-07-01   &   S2-W2-E2   &   K   &   43      &   5       &   \object{HD 165920}, \object{HD 163955}   &   \nodata \\
    &   2012-06-27   &   W1-W2-E2   &   K   &   28      &   4       &   \object{HD 163955}   &   \nodata \\
    &   2012-06-28   &   S1-E1-E2   &   K   &   35      &   5       &   \object{HD 159743}, \object{HD 170622}, \object{HD 163955}   &   \nodata \\
    &   2012-06-29   &   W1-E1-E2   &   K   &   42      &   6       &   \object{HD 159743}, \object{HD 163955}   &   \nodata \\
    &   2012-06-30   &   S1-S2-E1   &   K   &   14      &   2       &   \object{HD 163955}   &   \nodata \\
    &   2012-07-01   &   S1-S2-W1   &   K   &   14      &   2       &   \object{HD 159743}, \object{HD 163955}   &   \nodata \\
    &   2013-06-12   &   S1-W1-W2   &   K   &   21      &   3       &   \object{HD 163955}, \object{HD 170622}   &   \nodata \\
    &   2013-06-16   &   W1-W2-E2   &   K   &   14      &   2       &   \object{HD 170657}   &   \nodata \\
    &   2014-06-09   &   W1-E1-E2   &   H   &   23      &   3       &   \object{HD 163955}, \object{HD 164031}   &   \nodata \\
    &   2014-06-11   &   S2-E2      &   H   &   2       &   0       &   \object{HD 150366}   &   \nodata \\
    &   2014-06-12   &   S2-W1-E2   &   H   &   7       &   1       &   \object{HD 164031}   &   \nodata \\ \hline
    HD~190073   &   2010-09-29   &   S2-W2-E2   &    K    &   17   &   2   &   \object{HD 183936}   &   \nodata \\
    &   2011-06-15   &   W1-W2-E1   &    K    &   21   &   3   &   \object{HD 188385}, \object{HD 189509}   &   \nodata \\
    &   2011-06-18   &   W1-W2-E1   &    K    &   23   &   2   &   \object{HD 189509}, \object{HD 188385}   &   \nodata \\
    &   2011-06-20   &   W1-W2-E1   &    K    &   27   &   3   &   \object{HD 188385}, \object{HD 189509}   &   \nodata \\
    &   2011-06-23   &   S1-W1-W2   &    K    &   35   &   5   &   \object{HD 189509}, \object{HD 188385}, \object{HD 190498}   &   \nodata \\
    &   2011-06-24   &   S1-W2-E2   &    K    &   49   &   7   &   \object{HD 189509}, \object{HD 188385}   &   \nodata \\
    &   2011-06-25   &   S1-W2-E2   &    K    &   7    &   1   &   \object{HD 189509}   &   \nodata \\
    &   2011-06-26   &   S1-W1-W2   &    K    &   28   &   4   &   \object{HD 188385}, \object{HD 189509}, \object{HD 190498}   &   \nodata \\
    &   2011-06-27   &   S1-W1-E2   &    K    &   21   &   3   &   \object{HD 189509}, \object{HD 188385}, \object{HD 190498}   &   \nodata \\
    &   2011-06-28   &   S1-W2-E1   &    K    &   21   &   3   &   \object{HD 189509}, \object{HD 188385}   &   \nodata \\
    &   2011-07-01   &   S2-W2      &    K    &   2    &   0   &   \object{HD 188385}, \object{HD 189509}   &   \nodata \\
    &   2011-08-03   &   S1-W1-E2   &    K    &   9    &   1   &   \object{HD 189509}, \object{HD 188385}   &   \nodata \\
    &   2012-06-27   &   W1-W2-E2   &    K    &   35   &   5   &   \object{HD 188385}   &   \nodata \\
    &   2012-06-28   &   S1-E1-E2   &    K    &   42   &   6   &   \object{HD 189509}, \object{HD 188385}   &   \nodata \\
    &   2012-06-29   &   W1-E1-E2   &    K    &   35   &   5   &   \object{HD 189509}, \object{HD 188385}   &   \nodata \\
    &   2012-06-30   &   S1-S2-E1   &    K    &   49   &   7   &   \object{HD 189509}, \object{HD 188385}   &   \nodata \\
    &   2012-07-01   &   S1-S2-W1   &    K    &   21   &   3   &   \object{HD 189509}, \object{HD 188385}   &   \nodata \\
    &   2013-06-09   &   W1-W2      &    K    &   3    &   0   &   \object{HD 188385}   &   \nodata \\
    &   2013-06-10   &   W1-W2      &    K    &   21   &   0   &   \object{HD 188385}, \object{HD 190498}   &   \nodata \\
    &   2013-06-12   &   S1-W1-W2   &    K    &   7    &   1   &   \object{HD 188385}   &   \nodata \\
    &   2013-06-16   &   W1-W2-E2   &    K    &   14   &   2   &   \object{HD 188385}   &   \nodata \\
    &   2014-05-29   &   W2-E2      &    H    &   12   &   0   &   \object{HD 188385}, \object{HD 189712}   &   \nodata \\
    &   2014-05-30   &   S2-W2-E2   &    H    &   24   &   12   &   \object{HD 188385}, \object{HD 189712}   &   \nodata \\
    &   2014-06-02   &   S2-W1-E1   &    H    &   11   &   1   &   \object{HD 188385}, \object{HD 189712}   &   \nodata \\
    &   2014-06-03   &   S2-W1-E1   &    H    &   16   &   2   &   \object{HD 188385}, \object{HD 189712}   &   \nodata \\
    &   2014-06-04   &   S2-W1-E2   &    H    &   23   &   12   &   \object{HD 188385}, \object{HD 189712}   &   \nodata \\
    &   2014-06-05   &   S2-W1-W2   &    H    &   7    &   4    &   \object{HD 188385}, \object{HD 189712}   &   \nodata \\
    &   2014-06-07   &   S2-W1-W2   &    H    &   18   &   8    &   \object{HD 188385}   &   \nodata \\
    &   2014-06-11   &   E1-E2      &    H    &   4    &   0    &   \object{HD 188385}   &   \nodata \\
    \enddata
    \tablenotetext{}{The ``n.\@ $V^2$'' column indicates the number of squared visibility measurements collected per the particular observation and the ``n.\@ $T^3$'' column indicates the number of closure phase measurements collected.}
    \label{tab:climb}
\end{deluxetable}

The data from both Classic and CLIMB instruments were reduced using a pipeline developed in-house at the University of Michigan (available from the authors upon request).  The pipeline is designed specifically for robust selection of fringes of faint/low-visibility objects such as YSOs.  Squared visibilities are estimated using the power spectrum method and spectral windows are user-selectable based on the changing seeing conditions. For CLIMB, the closure phases were averaged using short, coherent time blocks across the fringe envelopes weighted by the magnitude of the triple-amplitude. The transfer function was monitored with regular observations of calibrators stars chosen to be single and nearly unresolved by the interferometer, selected with the JMMC \texttt{SearchCal} tool \rev{(\citealt{Chelli_2016})}. The pipeline was validated end-to-end using the known binary \object{WR~140} (\citealt{Monnier_2011}), including the sign of the closure phase.

Due to occasional sudden changes in sky or instrument conditions, the estimated calibration transfer function can be incorrect.  While we are working on objective, automated procedures based on signal-blind diagnostics, we currently rely on detecting outliers based on sudden, unphysical changes in our data as a function of time or baseline coverage.  While subjective, our outlier removal eliminated only \SI{3.6}{\percent} of the CHARA/Classic data and \SI{2.1}{\percent} of the CHARA/CLIMB data, and should not strongly affect our fitting results. Lastly, a minimum error floor is applied to the data from the pipeline to account for expected errors in our transfer function estimation mentioned above.  Based on earlier studies \citep{Tannirkulam_2008,Monnier_2011} and a new study from CLIMB using WR~140, we apply a minimum squared visibility errors of \SI{10}{\percent} relative error or additive \num{0.02} error, whichever is largest.  Generally for high visibilities, the relative error floor dominates the error budget while the additive error comes into play for long-baseline low-visibility points.

\subsection{VLTI  Interferometric Data}

We augment our CHARA observations with squared visibility and closure phase measurements obtained at the Very Large Telescope Interferometer (VLTI) at Cerro Paranal, Chile.  These data greatly increased the $(u, v)$ coverage of our data set at intermediate-length baselines of up to \num{128} meters.

We made use of archival data of HD~163296 and HD~190073 collected with \rev{the} AMBER instrument (\citealt{Petrov_2007}) spanning multiple nights in 2007-08 and in 2012, respectively (see Table \ref{tab:amber}). These
measurements \rev{were recorded} in low spectral resolution mode ($R = 30$) \rev{spanning H- and K-bands}.  The data were reduced using \rev{\texttt{amdlib} (v3.0.9; http://www.jmmc.fr/data\_processing\_amber.htm)} software where we derived
K-band visibilities
and closure phases. During the data reduction, only data with the \SI{20}{\percent} best fringe \rev{signal-to-noise ratio (}SNR\rev{)} were taken into account \citep{Tatulli_2007}. Because the observations were conducted without a fringe tracker, the data showed a typical spread of the optical path difference (OPD) between \SIrange[range-units = single, range-phrase = --]{20}{60}{\micro\meter}. To minimize the effect of the OPD spread and the resulting influence of atmospheric degradation on the calibrated visibilities, we used the histogram equalization method introduced by \cite{Kreplin_2012}.
Even with these corrections, the majority of the \rev{AMBER data spanning} H-band
\rev{remained of low quality due to poor calibration.}
Since we had access to
\rev{superior H-band} PIONIER data for both targets \rev{(see next paragraph)}, we excluded the AMBER H-band data \rev{in the present analysis},
\rev{and kept} only the K-band data
\rev{here}.  To maintain congruity with the CHARA observations, AMBER data with spectral measurements in the range of \SIrange[range-units = single, range-phrase = --]{1.995}{2.385}{\micro\meter} were collapsed to a single point by averaging the individual squared visibility values to simulate a broadband measurement (adopting the median wavelength of each spectral measurement set as the effective wavelength).  The squared visibility errors on the collapsed measurements were taken to be

\begin{equation}\label{eq:error}
    \sigma_{\mathcal{V}^2} = \frac{\sum_i^N \Delta \mathcal{V}_i^2}{N\sqrt{N-1}},
\end{equation}

\noindent with $N$ referring to the total number of measurements in each collapsed data point (typically 7) and $\Delta \mathcal{V}_i^2$ the pipeline estimated errors on the individual measurements.  We then applied a minimum error floor of \SI{10}{\percent} relative error or additive \num{0.02} error, whichever is largest, adopted to be consistent with the errors chosen for the CHARA experiments.  Finally, we applied our outlier removal, in accord with the CHARA data procedure, which resulted in fewer than \SI{3}{\percent} of the data rejected.

\begin{deluxetable}{lcrlccl}
    \tablecaption{VLTI/AMBER Observations}
    \tablehead{\colhead{Target} & \colhead{Ref.} & \colhead{Date} & \colhead{Configuration} & \colhead{Band} & \colhead{n.\,sets} & \colhead{Calibrator(s)}}
    \startdata
    \object{HD~163296}   &   a   &   2007-08-31   &   H0-E0-G0      &   K   &   1   &   \object{HD 172051} \\
    &   a   &   2007-09-03   &   H0-E0-G0      &   K   &   2   &   \object{HD 139663} \\
    &   a   &   2007-09-06   &   D0-H0-G1      &   K   &   1   &   \object{HD 172051} \\
    &   b   &   2008-05-25   &   D0-H0-A0      &   K   &   8   &   \object{HD 156897}, \object{HD 164031} \\
    &   b   &   2008-05-27   &   D0-H0-G1      &   K   &   1   &   \object{HD 156897} \\
    &   b   &   2008-06-04   &   H0-E0-G0      &   K   &   2   &   \object{HD 106248}, \object{HD 156897} \\
    &   b   &   2008-06-05   &   H0-E0-G0      &   K   &   5   &   \object{HD 156897}, \object{HD 163955} \\
    &   c   &   2008-06-24   &   UT1-UT2-UT4   &   K   &   1   &   \object{HD 156897} \\
    &   d   &   2008-07-07   &   D0-H0-G1      &   K   &   5   &   \object{HD 160915} \\
    &   d   &   2008-07-09   &   K0-A0-G1      &   K   &   4   &   \object{HD 160915} \\
    &   e   &   2008-07-19   &   K0-A0-G1      &   K   &   1   &   \object{HD 156897} \\ \hline
    \object{HD~190073}   &   f   &   2012-06-07   &   UT2-UT3-UT4   &   K   &   4   &   \object{HD 179758}, \object{HD 188385} \\
    \enddata
    \tablerefs{\textsuperscript{a}Dataset 60.A-9054; \textsuperscript{b}Dataset 081.C-0794; \textsuperscript{c}Dataset 081.C-0851; \textsuperscript{d}Dataset 081.C-0124; \textsuperscript{e}Dataset 081.C-0098; \textsuperscript{f}Dataset 089.C-0130.}
    \label{tab:amber}
\end{deluxetable}

Additionally, we include in our data analysis recently published squared visibility and closure phase measurements from the PIONIER instrument (\citealt{Le_Bouquin_2011}) in H band, conducted by \cite{Lazareff_2017}; see observations and reduction summary therein.  Since these data were conducted in a low spectral resolution mode ($R = 15$), we collapsed the measurements to a single value in H band by averaging $V^2$ measurements (in like manner as we did for the AMBER instrument data in K band).  The PIONIER data consisted of 3 data points in the range \SIrange[range-units = single, range-phrase = --]{1.55}{1.80}{\micro\meter}.
\rev{We estimated the squared visibility errors by applying Equation \ref{eq:error} and then imposed a relative error floor of \SI{5}{\percent} to each measurement to account for visibility variations within H-band in each measurement.  No minimum additive error was applyed.  While these error floors are smaller than those applied to the Classic, CLIMB, and AMBER datasets, the overall data quality produced by the more recent PIONIER instrument is higher, as it employs fringe scanning mode that is better able to correct for seeing effects than AMBER and the Classic/CLIMB instruments.}

\subsection{Keck Aperture Masking Observations}

Finally, we include in our analysis archival short-baseline measurements of HD~163296 from the Keck aperture masking experiment in both H and K bands, collected on September 29, 1998. These short baseline data constrain the fraction of the total light distributed in a large-scale ``halo'' and also could detect stellar companions, if present.  Detailed descriptions of the experiment and the data analysis method were published by \cite{Tuthill_2000}. \rev{No additional percentage or \revtwo{additive} error floors were applied to the Keck data, which is dominated by systematic errors; all estimated $\mathcal{V}^2$ errors exceed \SI{12}{\percent} which is larger than any estimated seeing variations.}  Unfortunately, \rev{Keck aperture masking} observations of HD~190073 were not conducted.

Note that the Keck masking experiment has a roughly 5 times larger field-of-view (FOV FWHM $\sim$\ang{;;1}/\ang{;;1.3} for H/K bands) for ``incoherent flux'' than the single-mode fiber combiner PIONIER (FOV FWHM $\sim$\ang{;;0.18} for H band) and similar field of view to the CHARA combiners (seeing limited FOV FWHM $\sim$\ang{;;1}--\ang{;;2}) -- light beyond this FOV does not get included in the calibration of the visibility.

\section{Data}
\subsection{Squared Visibilities}

\begin{figure}[htbp]
    \begin{center}
    \includegraphics{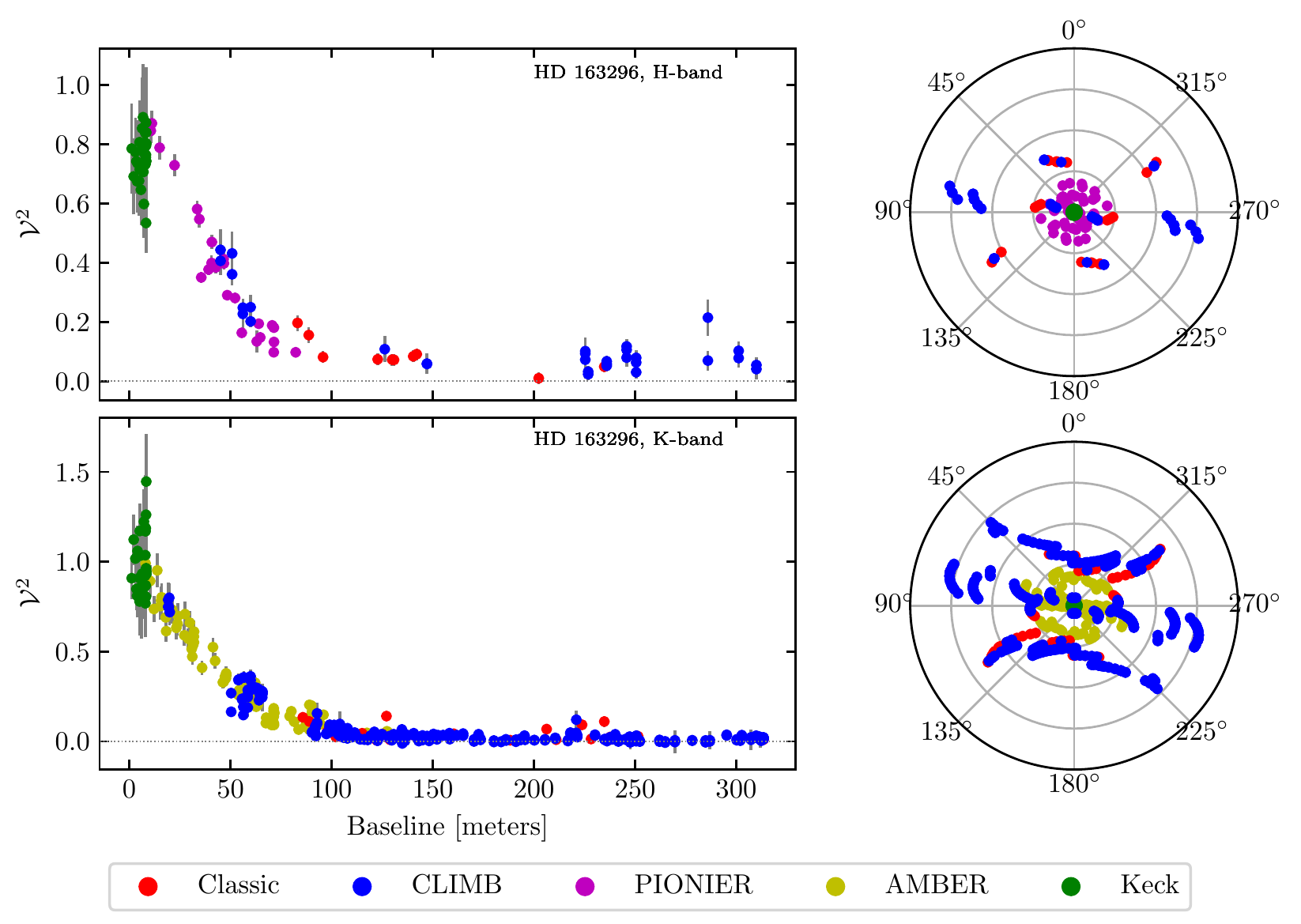}
    \caption{H and K band squared visibilities measurements of HD~163296 shown on the left and corresponding $(u, v)$ coverage on the right.  New data presented here from Classic/CLIMB data sets (Tables \ref{tab:classic} and \ref{tab:climb}) and from the AMBER experiment (Table \ref{tab:amber}).  For the $(u,v)$ plots, $u$ increases to the left (east), $v$ increases to the top (north) and the concentric axis circles are drawn with increasing radii in units of \SI{100}{\meter}.}
    \label{fig:mwc275raw}
    \end{center}
\end{figure}

\begin{figure}[htbp]
    \begin{center}
    \includegraphics{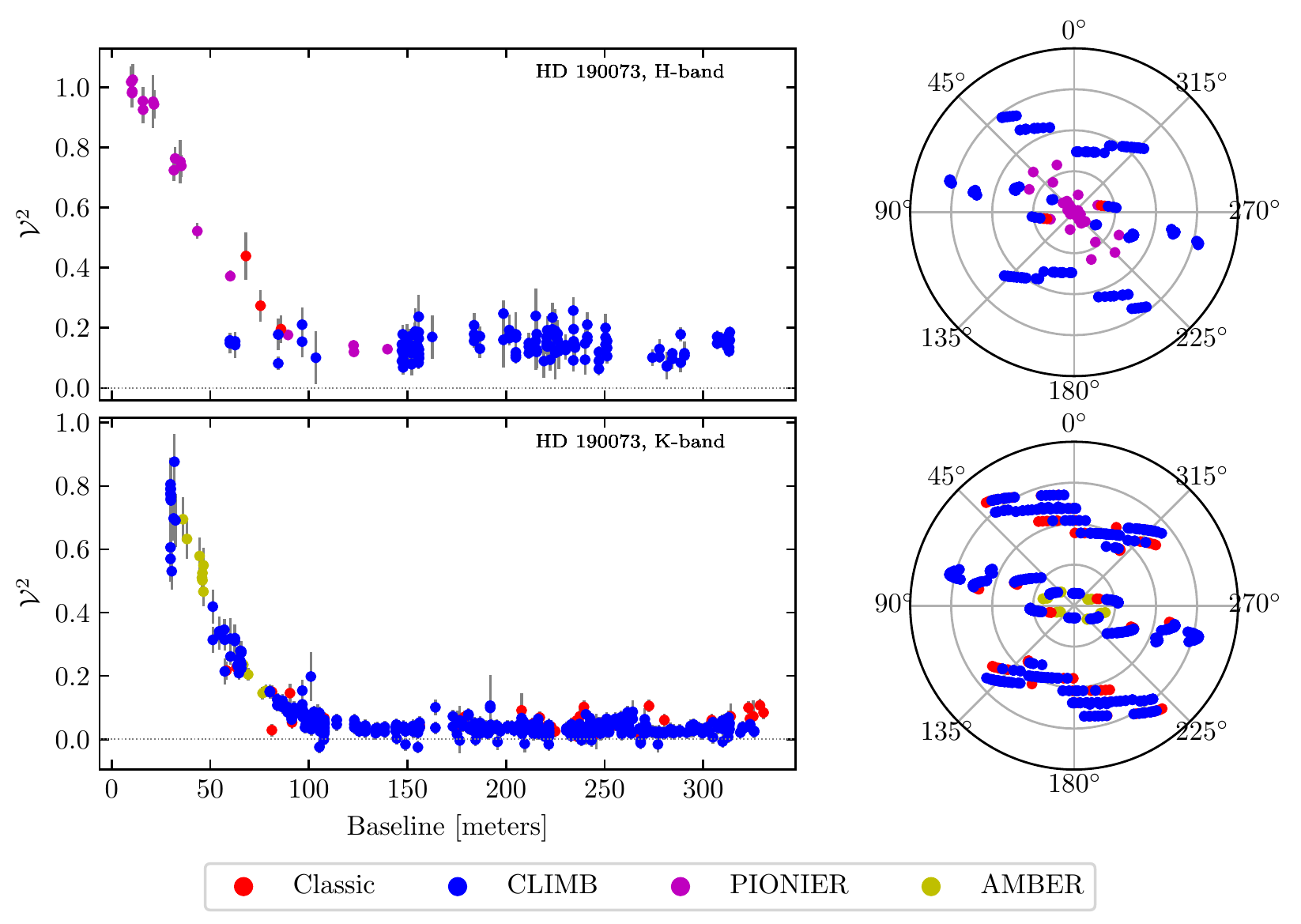}
    \caption{H and K band squared visibility measurements HD~190073.  New data presented here from Classic/CLIMB data sets (Tables \ref{tab:classic} and \ref{tab:climb}) and from the AMBER experiment (Table \ref{tab:amber}).}   
    \label{fig:v1295aqlraw}
    \end{center}
\end{figure}

The combined squared visibility measurements and full $(u,v)$ coverage from all instruments are shown in Figure \ref{fig:mwc275raw} for HD~163296 and in Figure \ref{fig:v1295aqlraw} for HD~190073.  
The $(u,v)$ coverages for both objects are rather well sampled in K band, providing a sufficient number of points at all covered baselines to infer a good picture of the radial brightness distribution at sub-\si{\au} scales.   Nonetheless, there are notable gaps in the data.  HD~163296 does not have excellent coverage at long baselines along the north-south direction, but is well sampled otherwise.  HD~190073 is missing intermediate baseline information in the north-south direction as well as intermediate/long baseline measurements along the east-west direction.  Moreover, due to the lack of Keck aperture masking measurements for HD~190073, there are no K band $(u,v)$ points at baselines shorter than \SI{29.9}{\meter} which are necessary to constrain the emission contribution due to a  large scale, over-resolved, halo component in the system.  The $(u,v)$ coverage for both objects in H band is not as well sampled, particularly at long baselines.

We can learn important qualitative features about both objects by inspecting their visibility curves.  For HD~190073, there are sufficient short baseline measurements from PIONIER to show that any large scale halo component present in the system contributes negligibly to the total flux received from the system, as the short baseline points indicate that the squared visibility is consistent with unity at zero baseline, within instrumental uncertainties.

For both targets, at baseline separations exceeding \SI{120}{\meter}, the squared visibilities remain approximately constant, especially in the K-band data; the lack of discernible oscillations about the asymptotic value at long baselines strongly suggests that the inner emission lacks sharp boundaries where the emission intensity changes quickly over a small radius range, a result first hinted at by \citet{Tannirkulam_2008} for HD 163296 with a mere eight squared visibility measurements at long baselines.  In the H band, at first glance it appears that there may be some oscillation about the asymptotic value;  however, this is coupled with larger uncertainties than in K-band due to poorer seeing as well as a sparser $(u, v)$ coverage, indicating that the disk profile in H-band may nonetheless be consistent with a smooth distribution as it is in the K band.  Some high H-band visibilities for HD~190073 between \SIrange[range-units = single, range-phrase = --]{200}{300}{\meter} may be due to mis-calibration.  Additionally, although our observations do not appear to coincide with periods of flux variability, we can not rule out their occurrence within our dataset \citep{Ellerbroek_2014}.

\subsection{Closure Phases}\label{sec:closurephase}

\begin{figure}[htbp]
    \begin{center}
    \includegraphics{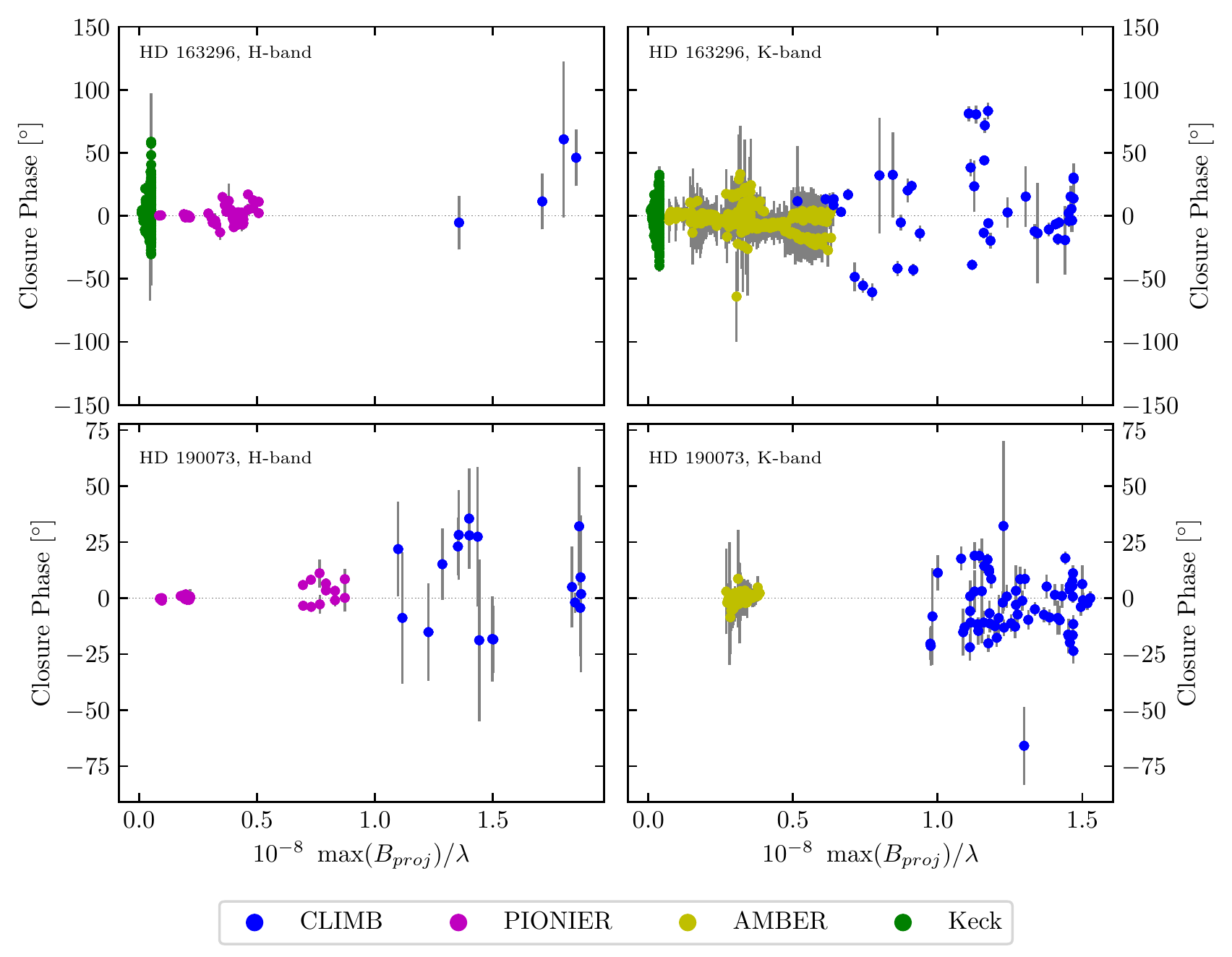}
    \caption{Top: measured closure phases of HD~163296.  Bottom: measured closure phases of HD~190073.}
    \label{fig:closurephase}
    \end{center}
\end{figure}

The closure phase measurements for both objects in H and K bands are presented in Figure \ref{fig:closurephase}.  Per convention, each closure phase measurement is plotted with respect to the longest baseline in the bispectrum closing triangle with the standard caveat that this is not necessarily the baseline along which the true Fourier phase is the strongest.  Note that the AMBER and PIONIER measurements plotted in the figure have not been ``collapsed'' as they have been the squared visibility case, instead the full spectral range of measurements are plotted individually within their respective bands.  To distinguish individual wavelength measurements collected at the same baseline, we plot spatial frequency rather than baseline distance along the abscissa.

We note that, for the most part, closure phase measurements with closing triangles on short to intermediate baselines are consistent with a value of zero, indicating that the brightness distribution is roughly point symmetric on larger spatial scales.  At longer baselines, we see a departure in the presumed point symmetry, indicating that fine scale structure is present, perhaps due to orbiting clumps in the inner disk.  It is intriguing that we see no clear evidence of this fine-scale structure in the visibility curve although the high data uncertainties\rev{, especially at long baselines where the SNR is on the order of 1-2,} likely mask the expected variations.

Finally, we note that the strongest closure phases, measured with the CHARA/CLIMB instrument, were conducted over a span of 4 years, during which time variability in the disk structure may have occurred.  Thus, the presented closure phase measurements cannot necessarily be considered as a representative snapshot of the disk asymmetries at any given time.

\section{Modeling}

\begin{deluxetable}{lCp{1.5in}}
    \tablecaption{Geometric Primitive Models}
    \tablehead{\colhead{Name} & \colhead{Visibility $V_{ext}(b)$} & \colhead{Comment}}
    \startdata
        Gaussian &  \displaystyle \exp\left[\frac{-(\pi\theta b)^2}{4\ln 2}\right] &  $\theta$ is the FWHM in \si{\milliarcsecond}. \\[16pt]
        Disk &  \displaystyle \frac{2 J_1(\pi\theta b)}{\pi\theta b} &  $\theta$ is the diameter in \si{\milliarcsecond}. \\[16pt]
        Thin Ring &  \displaystyle J_0(2\pi\theta b) &  $\theta$ is the diameter in \si{\milliarcsecond}. \\[16pt]
    \enddata
    \tablecomments{These models assume point symmetry where $b = \sqrt{u^2 + v^2}$ is the baseline in dimensionless units $B/\lambda_\text{eff}$.  The building blocks were taken from \cite{Berger_2007}}
    \label{tab:primitive}
\end{deluxetable}

We fit simple geometrical models to characterize the global properties of the extent and configuration of the inner-\si{\au} emission between our two targets.  Our focus is, primarily, to compare the spatial sizes of the emission seen in the H and K bands without trying to impose a particular interpretation on the data.  Therefore, we opt not to perform the sort of detailed radiative transfer modeling typically found in the current literature (eg.\@ \citealt{Aarnio_2017}), based on many as-of-yet unobservable chemical and geometrical properties of these systems.

While the new data presented in this paper provide the critical long baseline information necessary to constrain the sharpness of features in the inner disk to sub-\si{\au} resolution, the instrumental sensitivities of the squared visibility measurements are, unfortunately, not quite high enough to uniquely determine the true emission distribution.  We are however able to paint a crude general picture of the surface brightness distribution in the inner few \si{\au} of HD~163296 and HD~190073, which we hope will be able to provide a starting point for future radiative transfer modeling efforts.

\subsection{Model Building}\label{sec:modelbuild}

We tested three simple geometric emission models to fit to the squared visibility measurements, each composed of three components: a point source representing the unresolved star, an extended circumstellar emission component, and an over-resolved halo component.  These components provide fractional flux contributions $f_{ps}$, $f_{ext}$, and $f_{halo}$, respectively, such that they sum to unity.

A crucial initial step to our model analysis was to de-project the data into a face-on orientation.  To do this, our data were rotated and stretched onto an ``effective baseline''
%
%
by transforming the $(u, v)$ coordinates of each measurement into a $(u', v')$ frame via

\rev{
\begin{equation}\label{uvprime}
    \left(\begin{array}{c} u' \\ v' \end{array}\right) =
    \left(\begin{array}{cc}\sin P\!A & \cos P\!A \\ -\cos P\!A \cos i & \sin P\!A \cos i \end{array}\right)
    \left(\begin{array}{c} u \\ v \end{array}\right)
\end{equation}
}

\noindent where $i$ and $P\!A$ are the inclination and position angles of the disk, respectively.  These quantities were left as free parameters for out fitting routine.

With the data correctly de-rotated, the functional form of the total squared visibility measured at an effective baseline length $b$ is given by

\begin{equation}
    V^2(b) = \left[f_{ps} + f_{ext} V_{ext}(b)\right]^2,
\end{equation}

\noindent where $V_{ext}(b)$ is a ``primitive'' function describing an extended emission component from Table \ref{tab:primitive} and the visibilities of the star itself provide a constant offset contribution since the diameters of the stars themselves are unresolved at the baselines we probe.  Note that the halo flux contribution is not explicitly expressed in the equation; the addition of this component is to allow the model fits to have sub-unity visibility values at zero baseline.  The halo, being over-resolved, is dominant at baseline separations smaller than the shortest baseline measurements available and in reality raises the visibility to unity at the origin while making a negligible contribution at the baselines measured.

Our extended geometrical models were chosen to distinguish between two general scenarios of the inner disk emission: either the emission is largely constrained to a thin ring, representing the illuminated inner rim of a truncated dust disk, or the emission emanates primarily from a region within this supposed sublimation radius.
In the latter case, we fit two possible models: one in which the emission has a uniform disk surface brightness and another in which the brightness distribution is Gaussian in form. \rev{In all cases, we allow the H- and K-band sizes to fit independently in order to best minimize their residuals; we note that this is the case even in our ``inner rim'' ring models where the H- and K-band sizes would physically be expected to be the same.}

We constructed all our models with point-symmetric brightness distributions.  Strictly speaking, such an assumption is only valid if the closure phase measurements are all consistent with \SI{0}{\degree} or \SI{180}{\degree}, which is not the case in the present data, especially for HD~163296 at longer baselines.  However, as we argued in \S\ref{sec:closurephase}, the closure phases likely indicate fine scale asymmetries in the overall brightness distribution.  Since our purpose in this paper is only to characterize the general large-scale emission profile's size and sharpness, choosing a point symmetric brightness distribution is justified considering that the closure phases measured along the primary lobe in visibility in both targets are consistent with zero.

\subsection{Model Fitting}

Each model was fit simultaneously to the H- and K-band data for each object.  Because the emission component was assumed to be coplanar, the inclination and position angles of the extended emission component were fixed together.  This is generally not desirable, but was a necessary constraint due to the relative overall lack of H-band squared visibility data.  Component flux contributions and size parameters were allowed to vary, but the sum of the point source and extended emission fluxes (or in other words, the expected halo contribution) were fixed together between the two bands, due to the lack of reliable data at short baselines between the different instruments, which is partially a consequence of their differing fields of view.  In the case of HD~190073, the data do not suggest the presence of a halo component.  In fact, since \rev{nominal} fits of the data for HD~190073 tended to slightly overshoot \rev{a value of 1.0} at zero baseline \rev{(a nonphysical result), we added an additional constraint that the overall summed total fractional flux contributions of the point source and extended emission equaled unity there for this object, effectively eliminating any halo contribution.}

Our fits employed the Levenberg-Marquadt gradient descent algorithm to minimize the $\ell$1-norm statistic.  To ensure we found the best fit, we initialized each model with \num{200} random sets of starting parameter values, and selected the fit with the overall lowest $\ell$1-norm value.  This statistic was selected over the more typical $\ell$2-norm since our squared visibility error estimates are non-Gaussian due to seeing and calibration effects and the underlying covariance is generally poorly understood.  Moreover, there are several obvious outliers in the data set, such as squared visibility measurements with negative values due to bias corrections, to which the $\ell$1-norm statistic is more robust.  That said, the final fit parameters were not found to be significantly different when the fits were conducted with the $\ell$2-norm.

The resulting parameter values and reduced $\ell$1-norms of our model fits for HD~163296 and HD~190073 are presented in Tables \ref{tab:mwcres} and \ref{tab:v12res}, respectively.  We estimated the uncertainties on the fit parameters by performing \num{1000} sets of bootstraps on the individual data points with \num{10} random sets of starting values in the neighborhood (of a few $\sigma$) of each of the best fit values using the same fitting procedure described above.  Reported uncertainties in the fit parameters were found by locating the most compact \SI{68}{\percent} in the bootstraps and computing two sided errors around the best fit value of all the data.  For all parameters except the inclination and position angles, the two sided errors were nearly symmetric.  It should be noted however that the stated errors are likely under-estimating the true uncertainties due to our neglect of possible correlated errors in our datasets expected due to seeing variations.  Another concern is that incomplete (and in some locations, especially in H-band, very sparse) $(u,v)$ coverage causes certain data points to have an artificially large weight on the final fit result, amplifying possible systematic uncertainties.

\begin{figure}[htbp]
    \begin{center}
    \includegraphics{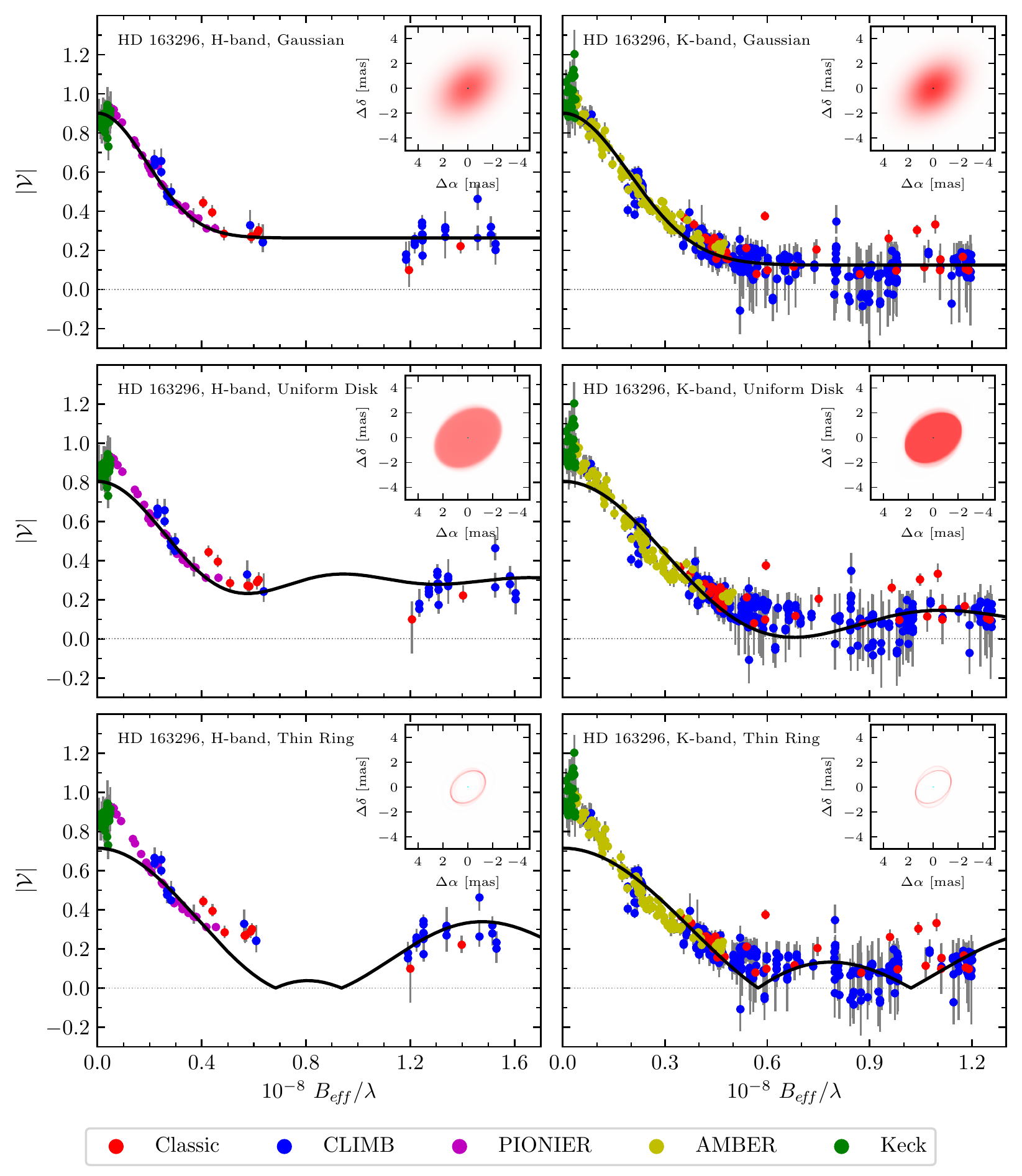}
    \caption{Model fit results of HD~163296, with H band fits on the left and K band fits on the right.  Comparison of the different models tested is shown vertically.  Inset in each plot is a stacked image of the emission distribution generated by the individual bootstrap fit parameters.}
    \label{fig:mwcres}
    \end{center}
\end{figure}

\begin{figure}[htbp]
    \begin{center}
    \includegraphics{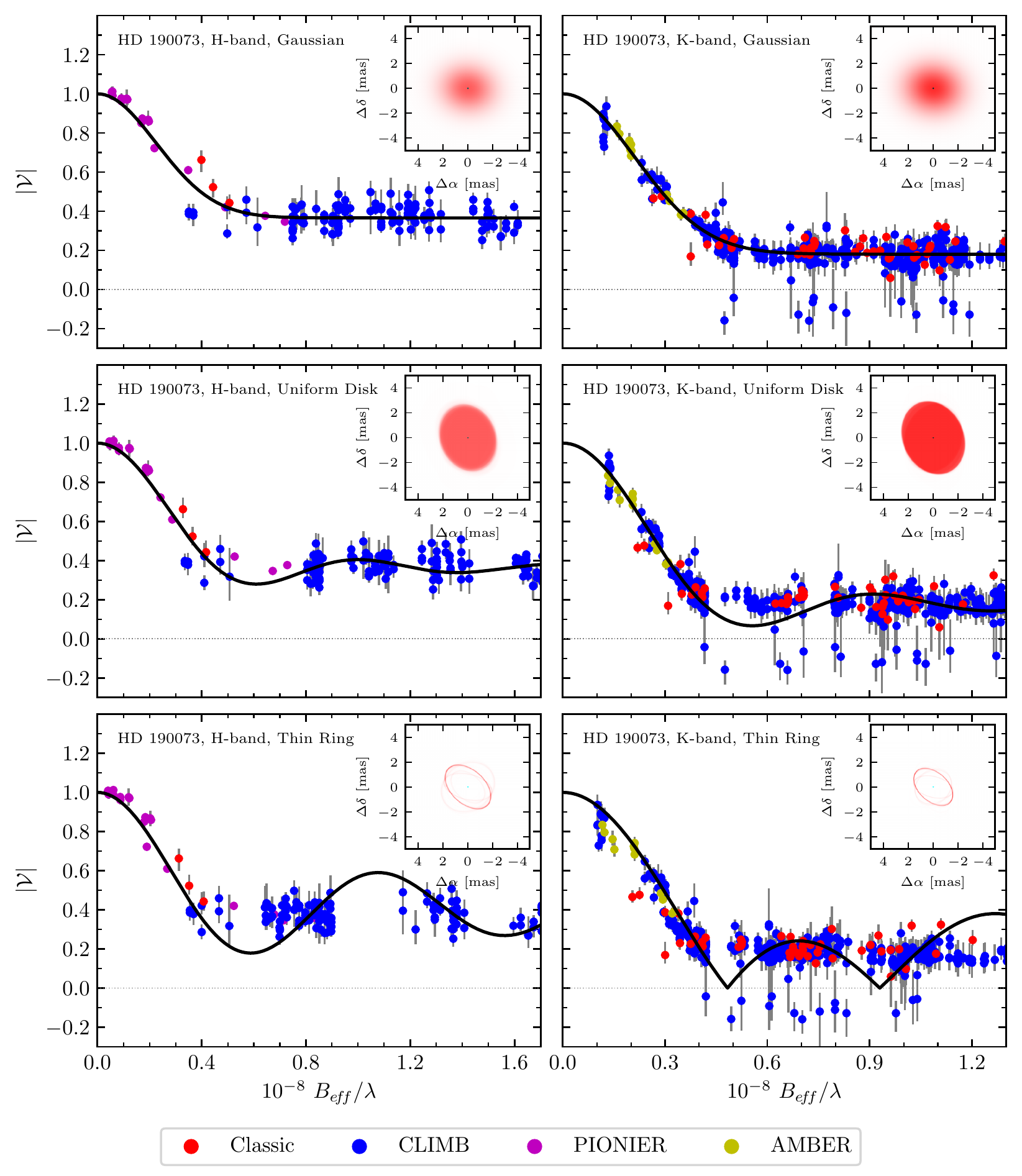}
    \caption{Model fit results of HD~190073.  See the caption in Figure \ref{fig:mwcres} for details of the layout.}
    \label{fig:v12res}
    \end{center}
\end{figure}

In Figures \ref{fig:mwcres} and \ref{fig:v12res}, we overplot the best fit models on the de-projected interferometric measurements data with the modulus of the visibility along the ordinate and the effective baselines along the abscissa.  While the fits themselves were conducted against the actual squared visibility measurements, visibility provides a better sense of the contribution importance of the various model components.  As per convention, measurements recorded with negative squared visibilities retain that sign in the absolute visibility plots as well.  We also inset stacked spatial image reconstructions for each of the \num{1000} bootstrap fit results in the aforementioned figures to give a visual sense of the presumed spatial distributions and range of uncertainties in our model fits.

\section{Results and Discussion}\label{sec:results}

We found that, of the simple models we tested, the Gaussian surface brightness distributions of the inner disk systematically outperformed the other models, visually and statistically, in describing the squared visibility measurements for both targets.  For HD~163296, where the visibility at zero baseline was allowed to fit as a free parameter, only the Gaussian model provided good agreement with the data at short baselines; the uniform disk and thin ring models significantly underestimated the visibility measurements.  A key feature of the Gaussian model is that the visibility quickly and monotonically asymptotes to the flux value of the star, without any ``ringing'' seen at longer baselines.  This description is consistent with the observed data, to within the degree of uncertainty assumed given the systematics of the instruments used.  We note that the apparent ringing seen in the H-band CLIMB long-baseline data may be due to poor calibration and sparse $(u, v)$ plane sampling rather than a real effect, as systematic errors are more prominent in H-band and the long-baseline K-band data are consistent with a flat visibility profile.  While monotonically decreasing behavior at long baselines is not unique to Gaussian models and also describe power-law-like models such as the Pseudo-Lorentzian profile described in \cite{Lazareff_2017}, such models disagree with observations at short baselines where the observed data indicate a concave down 
visibility profile whereas Pseudo-Lorentzian models are concave up. 

The uniform disk models we tried fit fairly well to the data based solely on the $\ell$1-norm values, though were overall not as successful at describing the data as the Gaussian models.  For instance, in the case of HD~163296 where the fit visibility at zero baseline was allowed to vary as a free parameter, the disk models do noticeably underestimate data at short baselines.  In K-band for both targets, the ringing at intermediate baselines predicted by the uniform disk models do not visually match the observations.  At H-band, there appears to be better convergence between the model and the data at long baselines, but as mentioned earlier, this may well be due to \rev{sparse} (u,v) coverage and instrumental systematic effects in the H-band calibration of the CLIMB data.

Finally, the thin ring models, which prior to the work of \cite{Tannirkulam_2008} were the favored picture of the inner emission, performed the worst during the fitting process.  In the case of HD~163296, the ring models dramatically underestimate the observed visibilities at short baselines.  For HD~190073, the chosen inclination and position angles of the ``best'' fit appear to merely exploit our data's under-sampled $(u, v)$ coverage, and even still produce $\ell$1-norm values in H and K band which are larger than for the other models tested, respectively. \rev{It is interesting to note that the fits in \cite{Lazareff_2017} favored a more ring-like emission profile for HD~190073 to that of the disk-like geometry we find in this work.  It is difficult to assess the root of this discrepancy, which may point to deficiencies in our models since the true distributions are likely not exact Gaussians or rings.  We do note that the discrepancies in fit position angles may play a role in the presence or \revtwo{absence} of \rev{oscillations} in $\mathcal{V}^2$ characteristic of a ring-like geometry.  Furthermore, HD~190073 is more \revtwo{susceptible} to errors in fitting the position angle as it appears to be more face-on than HD~163296.  We point out that the longer baseline CHARA data we examine in this work is better able to constrain the general geometry and orientation better than PIONIER data alone.  However, the overall $(u,v)$ sampling of the square visibility data presented in this work and especially in \cite{Lazareff_2017} contain large gaps in coverage, and are likely the most important factor in the discrepancies noted.  We point out that our results for HD~163296 are in good agreement with \cite{Lazareff_2017} indicating robustness between the methodologies employed in both works.}

\begin{deluxetable}{lcccccccl}
    \tablecaption{HD~163296 Results}
    \tablehead{\colhead{Model} & \colhead{Band} & \colhead{$i$ [\si{\degree}]} & \colhead{$P\!A$ [\si{\degree}]} & \colhead{$\Theta$ [\si{\milliarcsecond}]$^\tablenotemark{a}$} & \colhead{$f_{ext}$} & \colhead{$f_{ps}$} & \colhead{$\ell$1-norm/N} & \colhead{comment}}
    \startdata
\multirow{2}{*}{Gaussian} & H & \multirow{2}{*}{$45.0^{+2.8}_{-1.1}$} & \multirow{2}{*}{$131.3^{+1.9}_{-2.3}$} & \num{ 4.08 \pm 0.10} & \num{ 0.64 \pm 0.02} & \num{ 0.26 \pm 0.01} & 0.85 & \multirow{2}{1in}{Best fit} \\
 & K &  &  & \num{ 3.98 \pm 0.08} & \num{ 0.78 \pm 0.02} & \num{ 0.12 \pm 0.01} & 0.81& \\
\multirow{2}{*}{Uniform Disk} & H & \multirow{2}{*}{$43.7^{+2.3}_{-5.9}$} & \multirow{2}{*}{$124.5^{+4.7}_{-3.2}$} & \num{ 5.85 \pm 0.28} & \num{ 0.51 \pm 0.05} & \num{ 0.30 \pm 0.02} & 1.26 & \multirow{2}{1in}{Underestimates short baseline data} \\
 & K &  &  & \num{ 4.97 \pm 0.16} & \num{ 0.70 \pm 0.04} & \num{ 0.10 \pm 0.01} & 1.14& \\
\multirow{2}{*}{Thin Ring} & H & \multirow{2}{*}{$46.9^{+1.1}_{-6.6}$} & \multirow{2}{*}{$128.6^{+5.2}_{-8.0}$} & \num{ 3.12 \pm 0.12} & \num{ 0.54 \pm 0.07} & \num{ 0.18 \pm 0.05} & 1.98 & \multirow{2}{1in}{Underestimates short baseline data} \\
 & K &  &  & \num{ 3.20 \pm 0.28} & \num{ 0.60 \pm 0.06} & \num{ 0.11 \pm 0.02} & 1.25& \\
    \enddata
    \tablenotetext{a}{$\Theta$ indicates FWHM for the Gaussian model and the diameter for the uniform disk and ring models.  See \S\ref{sec:modelbuild} for definitions of the remainder of the parameters.}
    \label{tab:mwcres}
\end{deluxetable}

\begin{deluxetable}{lcccccccl}
    \tablecaption{HD~190073 Results}
    \tablehead{\colhead{Model} & \colhead{Band} & \colhead{$i$ [\si{\degree}]} & \colhead{$P\!A$ [\si{\degree}]} & \colhead{$\Theta$ [\si{\milliarcsecond}]\tablenotemark{a}} & \colhead{$f_{ext}$} & \colhead{$f_{ps}$} & \colhead{$\ell$1-norm/N} & \colhead{comment}}
    \startdata
\multirow{2}{*}{Gaussian} & H & \multirow{2}{*}{$32.2^{+5.0}_{-4.08}$} & \multirow{2}{*}{$81.6^{+5.6}_{-8.20}$} & \num{ 3.44 \pm 0.37} & \num{ 0.63 \pm 0.01} & \num{ 0.37 \pm 0.01} & 1.15 & \multirow{2}{1in}{Best fit} \\
 & K &  &  & \num{ 3.67 \pm 0.04} & \num{ 0.82 \pm 0.01} & \num{ 0.18 \pm 0.01} & 0.67& \\
\multirow{2}{*}{Uniform Disk} & H & \multirow{2}{*}{$36.8^{+1.9}_{-2.7}$} & \multirow{2}{*}{$22.6^{+5.9}_{-2.3}$} & \num{ 5.53 \pm 0.40} & \num{ 0.64 \pm 0.01} & \num{ 0.36 \pm 0.01} & 1.26 & \multirow{2}{1in}{Decent fit} \\
 & K &  &  & \num{ 6.07 \pm 0.11} & \num{ 0.82 \pm 0.01} & \num{ 0.18 \pm 0.01} & 0.90& \\
\multirow{2}{*}{Thin Ring} & H & \multirow{2}{*}{$50.9^{+1.5}_{-2.1}$} & \multirow{2}{*}{$48.0^{+6.3}_{-3.9}$} & \num{ 4.28 \pm 0.16} & \num{ 0.59 \pm 0.02} & \num{ 0.41 \pm 0.02} & 2.13 & \multirow{2}{1in}{Over-fits to poor $(u,v)$ coverage} \\
 & K &  &  & \num{ 3.63 \pm 0.14} & \num{ 0.88 \pm 0.01} & \num{ 0.12 \pm 0.01} & 1.51& \\
    \enddata
    \tablenotetext{a}{$\Theta$ indicates FWHM for the Gaussian model and the diameter for the uniform disk and ring models.  See \S\ref{sec:modelbuild} for definitions of the remainder of the parameters.}
    \label{tab:v12res}
\end{deluxetable}

\subsection{Comparison to Photometry}

\begin{figure}[htbp]
    \begin{center}
    \includegraphics{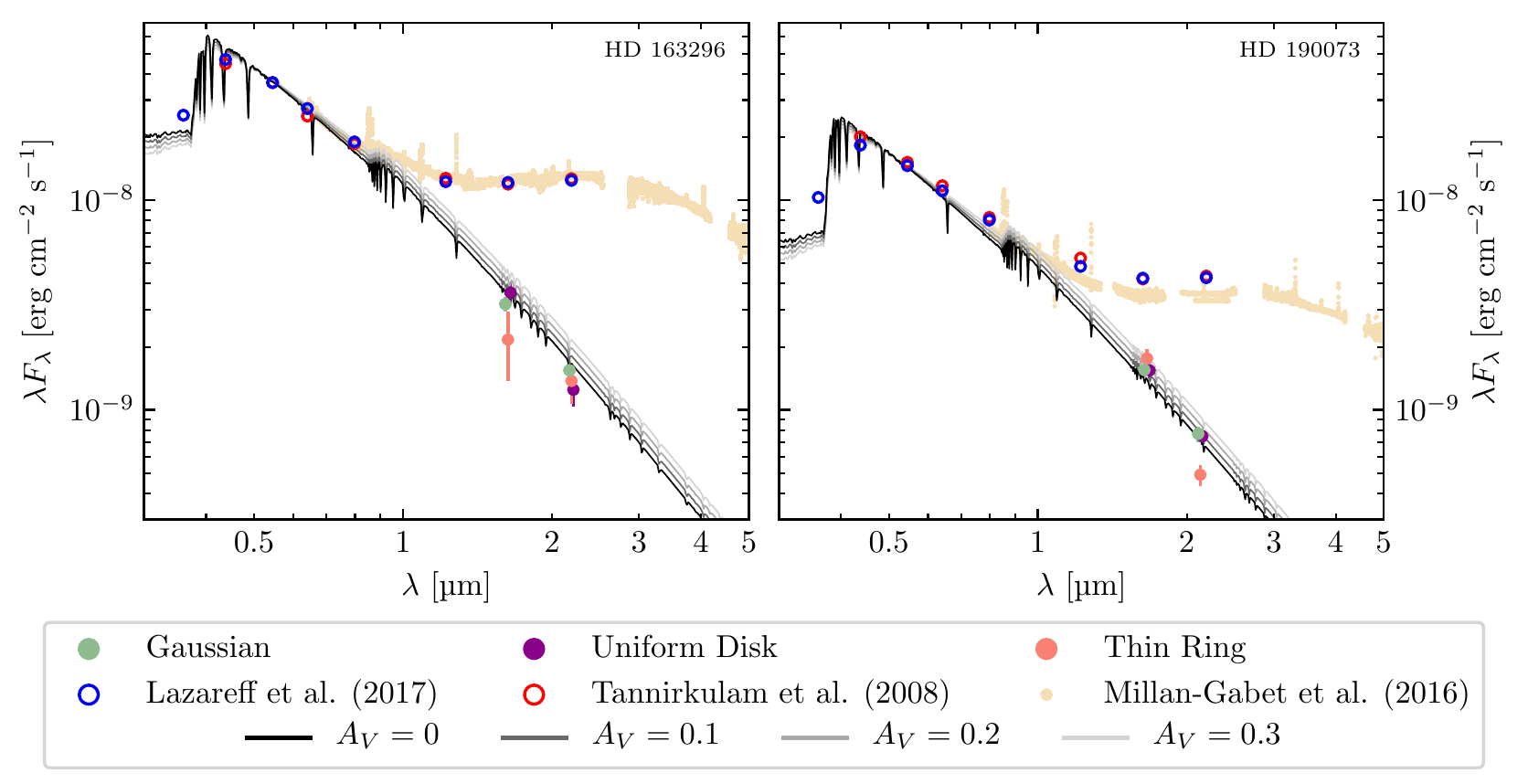}
    \caption{SEDs of our target objects and point source contributions at H and K bands as inferred by the interferometric models tested in this paper.  Note the interferometric points are slightly staggered for clarity.  Over-plotted on the data are model photospheres of \cite{Castelli_2004} \rev{at a range of extinctions fit to B, V, and R photometric observations by \cite{Lazareff_2017}.}  Left (HD~163296): \rev{Model photospheres with $T_{e\!f\!f} = \SI{9250}{\kelvin}$ and $\log g = 4.5$.} Right (HD~190073): \rev{Model photospheres with $T_{e\!f\!f} = \SI{9000}{\kelvin}$ and $\log g = 3.0$.}}
    \label{fig:sed}
    \end{center}
\end{figure}

In Figure \ref{fig:sed}, we use photometry collected by \cite{Tannirkulam_2008} in June, 2006, and by \cite{Lazareff_2017} in 2014, in conjunction with near-mid infrared SED measurements collected in June/July, 2007 for HD~190073 and in March, 2011 for HD~163296, first presented in \cite{MillanGabet_2016}, to construct crude SED profiles of the two objects spanning from near-ultraviolet to near-infrared wavelengths.  \rev{The long baseline interferometry allows us to estimate the flux contribution of the unresolved point source stellar photosphere in H and K bands by investigating the asymptotic value of the visibility oscillations.}
We note that any interstellar dust reddening should not significantly affect the H--K color at the distances to the stars, so fitting model photospheres to the interferometry data has the two-fold effect of testing the extended flux models and also determining the reddening to the source by extrapolating the NIR interferometry to measured visible fluxes.

\rev{We begin by \revtwo{modeling} the photosphere of HD~163296.  Due to it's proximity to earth (\SI{101.5}{\parsec}), the amount of interstellar dust extinction to the star is negligible (see \texttt{Bayestar17}; \citealt{Green_2018}).  Using measured B, V, and R band photometric points (\citealt{Lazareff_2017}), we fit tabulated model photospheres of \cite{Castelli_2004} with the closest available temperature, \SI{9250}{\kelvin}, and all $\log g$ values available (ranging from 1.5 to 5.0 in increments of 0.5).  We do not use the U band photometric point in the fit\revtwo{,} as it is expected to be amplified by accretion shocks\revtwo{, nor} points redder than R band\revtwo{,} as the disk contribution begins to dominate in the infrared.  We can report that unpublished visible light interferometry using the CHARA-VEGA instrument confirms that \SI{94 \pm 6}{\percent} of the R-band flux is coming from an unresolved source (private communication from Karine Perraut and Denis Mourard).}

\rev{The H- and K-band flux contributions to the resulting model stellar photosphere SED match the slope and has values within errors of the points estimated by the Gaussian disk model (see Figure~\ref{fig:sed}), further indicating this is a good approximation of the true emission profile.  From the photosphere fits, we infer the stellar luminosity and radius, and note that all model photospheres (with different $\log g$ values) obtain radii within \SI{3}{\percent}.  We compare these results to model isochrones generated by the \texttt{PARSEC} code (\citealt{Bressan_2012}) to predict a luminosity of \SI{16.1}{\solarluminosity}, a mass of \SI{1.9}{\solarmass} with $\log g = 4.3$, and an age of \SI{10.4}{\mega\year}.}

\rev{We repeat this process for HD 190073, using model photospheres with temperatures at \SI{9000}{\kelvin}.  Due to its greater distance (\SI{891}{\parsec}), we apply a reddening correction according to \cite{Clayton_1989}, assuming a typical $R_V = 3.1$.  We find the Gaussian fit interferometric points at H and K band are consistent with $A_V \sim 0.19$, which is in excellent agreement with the \texttt{Bayestar17} value of $A_V$ = \num{0.186 \pm 0.062}.  Primarily as a result of the new \textit{Gaia} DR2 distance estimate, we find that this star is incredibly young, at \SI{320}{\kilo\year}, and bright, with a luminosity of \SI{560}{\solarluminosity}.  Our fits also suggest this star is twice as massive as previously assumed, with $M = \SI{5.6}{\solarmass}$ and $\log g = 3.2$ (formerly \SI{2.84}{\solarmass}; \citealt{Catala_2007}).}

\begin{deluxetable}{lcc}
    \tablecaption{\rev{Adopted} Stellar Properties of Target Sources} 
    \tablehead{\colhead{Property} & \colhead{\object{HD~163296}} & \colhead{\object{HD~190073}}}
    \startdata
       $A_V$ & 0 & \rev{0.19} \\
       Mass & \rev{\SI{1.9}{\solarmass}} & \rev{\SI{5.6}{\solarmass}} \\
       Radius & \rev{\SI{1.6}{\solarradius}} & \rev{\SI{9.8}{\solarradius}} \\
       $T_\text{eff}$ & \rev{\SI{9250}{\kelvin}} & \rev{\SI{9000}{\kelvin}} \\
       Luminosity & \rev{\SI{16}{\solarluminosity}} & \rev{\SI{560}{\solarluminosity}} \\
       Age & \rev{\SI{10.4}{\mega\year}} & \rev{\SI{320}{\kilo\year}}
    \enddata
    \tablenotetext{}{}
    \label{tab:fitprop}
\end{deluxetable}

\subsection{Disk Orientation}

We note that the values we \revtwo{fit for} the disk inclination \revtwo{and} position angle \revtwo{of} HD~163296 were consistent with one another for all three models tested.  In particular, the values obtained with our Gaussian model fits ($i = \SI{45.0}{\degree}^{+\SI{2.8}{\degree}}_{-\SI{1.1}{\degree}}$, $P\!A = \SI{131.3}{\degree}^{+\SI{1.9}{\degree}}_{-\SI{2.3}{\degree}}$) agree well with values quoted in the literature of gas measured via millimeter interferometry (\citealt{Flaherty_2015}: $i=\SI{48.4}{\degree}$, $P\!A = \SI{132}{\degree}$) and dust from scattered light studies on large scales (\citealt{Monnier_2017}: $i=\SI{48}{\degree}$, $P\!A = \SI{136}{\degree}$).

The situation is not as clear for HD~190073.  While the Gaussian and uniform disk models found similar inclination angles, their fit position angles were incompatibly different.  The thin ring model fit also produced radically differing values in inclination and position angle to the other two models.  Unfortunately, there isn't much available in the literature to compare our derived values to.  \cite{Lazareff_2017} found values of $i = \SI{30 \pm 4}{\degree}$ and $\theta = \SI{159 \pm 3}{\degree}$ in their best fit model, however, these values are highly suspect due to a poorly sampled $(u, v)$ coverage in their \revtwo{dataset}.  Moreover, these data were used as part of the present analysis, and different results for the position angle were deduced.  Since large-scale scattered light images or mm-wave ALMA images of HD~190073 have not yet been obtained, there is not yet an independent measure with which to test consistency.  \cite{Fukagawa_2010} included HD~190073 in their survey sample of scattered light targets with the Subaru telescope, but obtained a null result in the poor seeing conditions where observation of the target was attempted.  

We remind the reader that we tied the H and K band disk inclination and position angles together due to our limited (u,v) coverage.  While the similar H band K band sizes for these objects suggest a common physical origin, there are mechanisms that might produce orthogonal position angles, such as dust caught in a disk wind. Future higher-quality data should investigate this possibility.

\subsection{Disk Sizes and Possible Emission Sources}\label{sec:emission}

Our Gaussian fit model for HD~163296 indicates that the inner disk is the same size in H and K bands, within uncertainty, with an H/K size ratio of \num{1.03 \pm 0.03}.  At a distance of \SI{101.5}{\parsec}, the 3.98~mas FWHM from our Gaussian model has a physical size of approximately \SI{0.41}{\au}.  For conventional large grains assuming our revised luminosity of \rev{\SI{16}{\solarluminosity}}, we expect dust at this distance $R\sim\SI{0.20}{\au}$ at an equilibrium temperature of \rev{$T\sim\SI{1750}{\kelvin}$} \citep[assuming dust wall backwarming,][]{Monnier_2005}.

In the case of HD~190073, the Gaussian fits indicate an H/K size ratio of \num{0.94 \pm 0.10}.  At a distance of \SI{891}{\parsec}, the 3.67~mas FWHM of our best-fitting Gaussian brightness profile in K band has a physical size of \SI{3.27}{\au}.  Using a revised luminosity of \rev{\SI{560}{\solarluminosity}}, we find dust at the distance $R\sim\SI{1.64}{\au}$ has an equilibrium temperature of \rev{$T\sim\SI{1500}{\kelvin}$}.

For both sources, the dust temperatures expected at the Gaussian FWHM location are close to the sublimation temperature for most dust species expected in the inner disks of YSOs.  
However, a significant fraction of the inner NIR emission appears to originate within the sublimation radius.
The current dataset analyzed here does not yield any definitive conclusions pertaining to the dominant source of the inner emission observed, however, some insights may be gleaned via the fit results.

We examine briefly a few possibilities:
\begin{enumerate}
\item{Conventional dust species might exist inside the expected dust evaporation radius due to shielding by optically-thick inner gas in the midplane.  Alternatively, some unidentified refractory dust species might exist that survive at $T>\SI{2000}{\kelvin}$.  Testing these hypotheses requires detailed radiative transfer \revtwo{modeling} of the gas distribution in the inner sub-AU and is beyond the scope of this paper, although the similar sizes for H and K band qualitatively support this scenario since we would not expect strong differences in the emission between H and K bands for these temperatures.}
\item{Transparent dust grains could exist close to the star without evaporating. For a giant star,  \citet{Norris_2012} inferred the  presence of iron-free silicates close to the stellar surface within the expected dust evaporation radius.
\rev{They} suggested that species such as forsterite (\chem{Mg_2SiO_4}) and enstatite (\chem{MgSiO_3}) are almost transparent at wavelengths of \SI{1}{\micro\meter} and we speculate that these grains might be found in YSO disks as well.  Observations of YSOs in polarized light might identify the same scattering signature as was done by \citet{Norris_2012} for mass-losing evolved stars.
}

\item{Hot ionized gas within \SI{0.1}{\au} could be optically-thick due to free-free/bound-free opacity, as seen for Be star disks (eg. \citealt{Sigut_2007}). Since A stars do not emit enough ionizing radiation to maintain a sufficient reservoir of ionized gas, there would need to be a local heating source in the midplane for this mechanism to be viable, possibly due to viscous or magnetic heating. This mechanism produces a sharply rising spectrum into the infrared and would have the largest impact on the differential H and K band sizes.  If such an optically-thick gas were located in a thin disk that extended all the way to the surface of the star, with an inclination of \SI{45}{\degree} in conjunction with the rest of the inner-disk, this could in principle block up to \SI{14}{\percent} of the central stars' NIR flux.}

\end{enumerate}
 
\section{Summary}

We combined broad-band infrared interferometric observations of \object{HD~163296} and \object{HD~190073} at H and K bands collected at CHARA, VLTI, and Keck to present the most complete $(u,v)$ sampled set of observations at the longest baselines available to date for these two objects.  These observations allow for us to examine the inner disk structures on milliarcsecond (sub-\si{\au}) scales.

We characterize these observations with simple point-symmetric geometric models to estimate the orientation, sharpness, and spatial emission distribution of the inner disk, fixing the inclination and position angle between H and K bands.  We focus on extracting general, qualitative features concerning the multi-wavelength emission geometry without a detailed radiative transfer analysis at this time.  Our models however are able to constrain the basic size and profile of the surface brightness distribution of in the few \si{\au} immediately surrounding the central pre-main-sequence star.

We find that a 2D Gaussian disk profile is best able to reproduce the squared visibility measurements collected for both targets in both H and K bands, producing superior fits than both uniform disk and ring models. If the inner disk is optically-thin, we confirm earlier indications that the bulk of the inner emission originates close in to the host star, well within the supposed dust sublimation radius inferred by SED modeling.  In conjunction with AB~Aurigae and HD~163296 previously studied by \cite{Tannirkulam_2008}, along with \object{MWC~614} studied by \citet{Kluska_2018}, \rev{and \object{HD~142666} studied by \cite{Davies_2018}}, HD~190073 is now the \rev{fifth} HAe object observed by CHARA with sufficient angular resolution to
rule-out a \rev{sharp, thin} ring-like geometry for the bulk of the NIR disk emission.

For both HD~163296 and HD~190073, we find small, near-zero closure phases for all baselines probing the main lobe of the disk emission (baselines $<$\SI{130}{\meter}) suggesting the large scale emission is point-symmetric.  That said, some triangles with longer baselines show significant non-zero closure phases that indicate asymmetries on the $<$\SI{2}{\milliarcsecond} scale.
We speculate that asymmetries or clumpy emission in the inner disk could explain this and should motivate future monitoring campaigns to see if these clumps exist and if they show orbital motion.

We speculate on the origin of the innermost disk emission that is closer to the star than expected.
One major new result here is that the H- and K-band sizes are nearly the same for both objects, within \SI{3 \pm 3}{\percent} for HD~163296 and within \SI{6 \pm 10}{\percent} for HD~190073. This points towards a single emission mechanism throughout the inner \si{\au} such as thermal dust emission, as opposed to a combination free-free gas emission close to star and thermal dust emission farther out as has been previously suggested. We also highlight the possibility that the inner-\si{\au} could be filled, at least partially, with glassy grains nearly transparent at UV and visible wavelengths, but scattering/emitting in the near-infrared.  With the advent of next generation polarimetric modes on upcoming instruments at CHARA, we will be able to test this hypothesis by measuring the inner emission in scattered light.

\rev{Finally, we use the \revtwo{extracted} point source contribution \revtwo{of} our Gaussian fit results along with \revtwo{measured} photometry to \revtwo{model} the photospheric contribution of our \revtwo{targets'} SEDs\revtwo{, allowing} us to estimate the age, mass, luminosity and radius of the central stars.  For HD~163296, we find that a photosphere with a temperature of \SI{9250}{\kelvin} provides a fit to our interferometric data and measured visible photometry assuming zero interstellar reddening.  For HD~190073, \revtwo{we find that} a photosphere \revtwo{at} \SI{9000}{\kelvin} observed through 0.19 magnitudes of visible extinction to \rev{matches} the interferometric and photometric data. \revtwo{Coupled with the new} \textit{Gaia} distance of \SI{891}{\parsec}, \revtwo{we find HD~190073 to be} a very luminous and massive young star with age $<$\SI{0.4}{\mega\year}.}

\acknowledgments
JDM and BRS acknowledge support from NSF-AST 1506540 and AA acknowledges support from NSF-AST 1311698.  CLD, AK, and SK acknowledge support from the ERC Starting Grant ``ImagePlanetFormDiscs'' (Grant Agreement No. 639889), STFC Rutherford fellowship/grant (ST/J004030/1, ST/K003445/1) and Philip Leverhulme Prize (PLP-2013-110).
FB acknowledges support from NSF-AST 1210972 and 1445935.
MS acknowledges support by the NASA Origins of Solar Systems grant NAG5-9475, and NASA Astrophysics Data Program contract NNH05CD30C.

We thank William Danchi and Peter Tuthill for use of the previously unpublished Keck aperture masking data and also Paul Boley for his Python OIFITS module (available at \url{https://github.com/pboley/oifits}).

The CHARA Array is supported by the National Science Foundation under Grant No. AST-1211929, AST-1636624, and AST-1715788.  Institutional support has been provided from the GSU College of Arts and Sciences and the GSU Office of the Vice President for Research and Economic Development.

This research has made use of the Jean-Marie Mariotti Center \texttt{SearchCal} (\citealt{Chelli_2016}) service (available at \url{http://www.jmmc.fr/searchcal}) co-developed by LAGRANGE and IPAG, and of CDS Astronomical Databases SIMBAD and VIZIER (available at \url{http://cdsweb.u-strasbg.fr/}).

This work has made use of data from the European Space Agency (ESA) mission
{\it Gaia} \citep{Gaia} (\url{https://www.cosmos.esa.int/gaia}), processed by the {\it Gaia}
Data Processing and Analysis Consortium (DPAC,
\url{https://www.cosmos.esa.int/web/gaia/dpac/consortium}). Funding for the DPAC
has been provided by national institutions, in particular the institutions
participating in the {\it Gaia} Multilateral Agreement.

This research has also made use of the SIMBAD database (\citealt{simbad}), operated at CDS, Strasbourg, France, and
the NASA's Astrophysics Data System Bibliographic Services.

\vspace{5mm}
\facilities{CHARA, VLTI, Keck}

\software{SearchCal, amdlib, OIFITS.py, Bayestar17, PARSEC}


\bibliographystyle{aasjournal}
\bibliography{bibliography}


\listofchanges

\end{document}